\documentclass[conference]{IEEEtran}
\IEEEoverridecommandlockouts
\usepackage[utf8]{inputenc}
\usepackage{cite}
\usepackage{amsmath,amssymb,amsfonts}
\usepackage{algorithmic}
\usepackage{graphicx}
\usepackage{textcomp}
\usepackage{xcolor}
\newtheorem{theorem}{Theorem}
\newtheorem{lemma}{Lemma}
\newtheorem{remark}{Remark}
\newtheorem{definition}{Definition}
\DeclareMathOperator*{\argmax}{arg\,max}

\def\BibTeX{{\rm B\kern-.05em{\sc i\kern-.025em b}\kern-.08em
    T\kern-.1667em\lower.7ex\hbox{E}\kern-.125emX}}
\begin{document}

\title{Stochastic Geometry-Based Modeling and Analysis of Beam Management in 5G}

\author{Sanket S. Kalamkar,$^{1}$ Fuad M. Abinader Jr.,$^{2}$ Fran\c{c}ois Baccelli,$^{1}$ Andrea S. Marcano Fani,$^{2}$\\ and Luis G. Uzeda Garcia$^{2}$\\
Affiliation: $^1$INRIA-ENS, Paris, France, $^2$Nokia Bell Labs, Paris, France
\thanks{E-mail: sanket.kalamkar@inria.fr, fuad.abinader@nokia-bell-labs.com.}
\thanks{This work has received funding from the European Research Council (ERC) under the European Union's Horizon 2020 research and innovation programme
grant agreement number 788851.}
}

\maketitle

\begin{abstract}
Beam management is central in the operation of dense 5G cellular networks. Focusing the energy radiated to mobile terminals (MTs) by increasing the number of beams per cell increases signal power and decreases interference, and has hence the potential to bring major improvements on area spectral efficiency (ASE). This benefit, however, comes with unavoidable overheads that increase with the number of beams and the MT speed. This paper proposes a first system-level stochastic geometry model encompassing major aspects of the beam management problem: frequencies, antennas, and propagation; physical layer, wireless links, and coding; network geometry, interference, and resource sharing; sensing, signaling, and mobility management. This model leads to a simple analytical expression for the effective ASE that the typical user gets in this context. This in turn allows one to find, for a wide variety of 5G network scenarios including millimeter wave (mmWave) and sub-6 GHz, the number of beams per cell that offers the best global trade-off between these benefits and costs. We finally provide numerical results that discuss the effects of different systemic trade-offs and performances of mmWave and sub-6 GHz 5G deployments.
\end{abstract}


\section{Introduction}



The ever-increasing demand in capacity for mobile communications makes it necessary to consider new implementation approaches/techniques that can significantly boost data rates and the area spectral efficiency (ASE) of mobile networks. One key enabler considered in 5G~\cite{5G_book} to face this demand is the use of the spectrum beyond the sub-6 GHz frequencies, known as millimeter wave (mmWave). More specifically, 5G relies on the spectrum above 20 GHz, where bandwidth of up to 400 MHz can be used to offer very high data rates (above 10 Gbps peak rates) and increase the network capacity; nevertheless, the sub-6 GHz bands, with up to 100 MHz of bandwidth, are still needed to ensure wide area coverage and data rates up to a few Gbps.

One of the key difficulties that mmWave frequencies face is their challenging propagation characteristics: they are subject to high path loss, penetration loss, and diffraction loss due to their millimetric wavelength, thus reaching short distances typically within a few hundred meters. But what was once considered a limitation makes nowadays mmWave a suitable candidate for small cells deployments, which can be used for network densification and capacity boosting.

To overcome the propagation challenges at mmWave frequencies, steerable arrays with a large number of antenna elements are used to create highly directional beams that concentrate the transmitted energy to achieve high gains and make the signal more robust to increase coverage. Moreover, mmWave communications must be designed to operate under mobility conditions, covering users in LOS (Line-of-Sight) and NLOS (Non-LOS) at pedestrian and vehicular speeds. This can become quite challenging since mmWave frequencies are highly sensitive to changes in the environment.
Thus, any mmWave-based system relies on beam management techniques 
to select the best beam during a base station (inter-cell) handover and quickly switch/reselect a new beam during intra-cell mobility to avoid beam misalignment and performance losses. Although more important for mmWave frequencies, beamforming techniques with highly directional narrow beams can also be used in sub-6 GHz frequencies to enhance the network performance. 

Nevertheless, beam-based communications come with implementation challenges, beam management procedures such as beam refinement and beam failure detection and recovery that introduce different signaling and delay overheads. Also, an efficient beam management relies on capturing systemic trade-offs that depend on the mobile terminal (MT) mobility, cell sizes, and the number/width of the beams. For instance, a large number of narrow beams improves the signal-to-interference-noise ratio (SINR), but it also leads to more frequent service interruption due to beam switching and beam misalignment, which degrade the network performance. Given such a complex nature of beam management, the system-level simulations, although crucial, are often time-consuming and expensive. Hence, to complement system-level simulations, in this paper, we aim to provide a tractable mathematical framework that permits a system-level analysis of beam management in 5G.

\subsection{Contributions}
\begin{enumerate}
    \item We provide a first mathematical optimization framework for the beam management problem in 5G. Using the well-established tools from stochastic geometry~\cite{FB_book, MH_book}, our proposed optimization yields the number of beams that maximizes the area spectral efficiency. 
    \item The optimization permits a system-level analysis in 5G. In particular, it takes into account the average speed of the MTs, mobility-induced beam-misalignment error, beam selections during base station handovers and beam reselections within a cell, and time overheads associated with these beam (re)selections. For instance, given an average speed of the  mix of MTs (pedestrians, bikes, and cars), the mix of geometries (some cells are bigger, other smaller, with the MTs either close or far from the serving base station), the mix for fading and blockages present in the network, our optimization gives the number of beams that provides the best ``sum rate" for MTs of all types in a large ball.
    \item While the proposed framework is generic and covers a variety of related problems, e.g., other beam-centric systems, we give numerical results to evaluate a 5G New Radio (NR)-compliant radio access network operating in a dense urban macro/picocell scenario, for both sub-6 GHz and mmWave frequencies. The results reveal the key inter-dependencies between network parameters and provide insights into the associated trade-offs.
\end{enumerate}

\subsection{Related Work}
The study of user handovers in cellular networks and their effect on various performance metrics such as throughput is a rich area~(see e.g., \cite{Survey_handover1, Survey_handover2, Tabassum} and the references therein). Most of the works have focused on only base station handovers (or, equivalently, cell handovers) in cellular networks~\cite{Tokuyama,Bartek,Teng,Arshad,Demarchou}. The work in~\cite{Li} is probably the closest one to our work. This work studies both initial beam selections during base station handovers and beam reselections within a cell and their associated overheads in a beam-centric mmWave cellular network. In particular, authors in \cite{Li} have obtained analytical expressions of inter-cell and inter-beam handover rates, based on which, the average spectral efficiency is calculated subject to overheads due to handovers. But, the work in~\cite{Li} considers only noise-limited scenario and ignores interference from other base stations, while our work considers both noise and interference in the analysis. Second, our work also studies the effect of handovers on optimizing the number of beams, while \cite{Li} assumes a fixed number of beams, i.e., no beam management. Third, unlike \cite{Li}, we consider blockages and beam misalignment due to mobility. 

Overall, there is a lack of understanding on how the MT mobility affects the beam management in the presence of interference, blockages, and beam misalignment error. 




\section{System Model}
\subsection{Network Model}
\begin{figure}
\centering
\includegraphics[scale=.48]{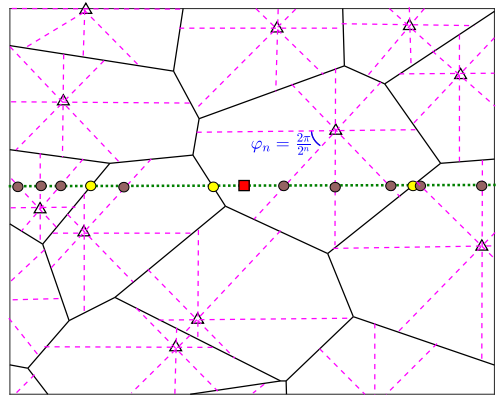}\vspace*{-5mm}
\caption{\small A snapshot of a Poisson cellular network with directional beamforming to an MT. In this case, each BS has $2^3= 8$ beams, i.e., $n = 3$. $\bigtriangleup$ : Base station (BS), red square : MT location, brown filled circle : Beam reselection location, yellow filled circle : BS handover location, solid black lines : Cell boundaries, dashed lines : Beam boundaries, and dotted line : Path of the MT.}
\label{fig:PVC}\vspace*{-2mm}
\end{figure}

As shown in Fig.~\ref{fig:PVC}, we consider a downlink cellular network, where the base station (BS) locations are modeled as a homogeneous Poisson point process (PPP) $\Phi \subset \mathbb{R}^2$ with intensity $\lambda$. We assume that the omnidirectional MT moves on a randomly oriented straight line with speed $v$. Without loss of generality, thanks to the isotropy and the stationarity of the PPP~\cite{FB_book}, this line of MT motion can be considered to be along the $X$-axis and passing through the origin. We assume that each BS always has an MT to serve. Also, a BS serves one MT at a time per resource block.  The MT associates itself with the nearest BS. Such an association results in BS cells forming a Poisson Voronoi tessellation as shown in Fig.~\ref{fig:PVC}. Let $X_0 \in \Phi$ denote the location of the serving BS of the MT at a given time. Without loss of generality, we can focus on the MT that is located at the origin at that time. After averaging over the PPP, this MT becomes the \textit{typical} MT.

\subsection{Beamforming Model}

A BS at $X \in \Phi$ uses beamforming to communicate with the typical MT. As shown in Fig.~\ref{fig:PVC}, we approximate the actual antenna pattern by a sectorized one, where each sector corresponds to one beam of the BS. For simplicity, we assume that each BS has $2^n$ beams with $n \in \mathbb{N}$, which corresponds to $2^{n-1}$ beam boundaries.\footnote{Here, a beam boundary is a line segment that connects two points on the cell boundary and passes through the location of the BS, as shown in Fig.~\ref{fig:PVC}.} Hence, the beamwidth is
\begin{align}
\varphi_n &= \frac{2\pi}{2^n} = \frac{\pi}{2^{n-1}}.
\label{eq:bw}
\end{align} 
We focus on a simple antenna pattern model where the main lobe is restricted to the beamwidth. Both the main and side lobe gains depend on the number $2^n$ of beams. In particular, the antenna gain $G_n$ is expressed as
\begin{align}
G_n(\psi) = \begin{cases}
               G_{{\rm m},  n}  & \text{if}~|\psi| \leq \varphi_n/2\\
               G_{{\rm s},  n} & \text{otherwise},
            \end{cases}
\end{align}
where $G_{{\rm m},  n}$ and $G_{{\rm s},  n}$ denote the antenna gains within the main and side lobes, respectively, as a function of the number $2^n$ of beams, and $\psi$ is the angle off the boresight direction. 

%

\subsection{Mobility-induced Beam Misalignment}

In 5G NR, the beam (re)selection occurs during a synchronization signal burst (SSB)~\cite{tut_bm}, which is done periodically with period $\tau$. During an SSB, the beam that has the MT within its beamwidth is selected for communication. We call this beam the \textit{reference beam}. Due to the mobility of the MT, it is possible that the MT has moved out of the beamwidth of the reference beam without selecting the new beam, and is still connected to the reference beam until the next SSB when the new reference beam is selected. Such a mobility-induced beam misalignment reduces the signal strength at the MT as it then lies within the side lobe of the reference beam chosen during a previous SSB.  At a given time, the beam misalignment probability $p_{\rm bm}$ depends on the speed $v$ of the MT, the duration $\tau$ between two SSBs, and the average distance between two beam reselections. Here, we propose a simple yet effective formula that captures the effects of these parameters on the probability of beam misalignment due to MT mobility in a snapshot of the network. We evaluate the probability $p_{\rm bm}$ that the MT is outside the beamwidth of the reference beam by:
\begin{align}
p_{\rm bm} = 1 - \exp\left(-\frac{v\tau}{1/\mu_{\rm s,b}}\right),
\label{eq:p_bm}
\end{align}
where $1/\mu_{\rm s,b}$ is the average distance between two consecutive beam reselections, with $\mu_{\rm s,b}$ being the linear intensity of beam reselection given in Theorem~\ref{thm:int_beam} in Section~\ref{sec:beam_resel}.
We can interpret \eqref{eq:p_bm} as
\begin{align*}
1- p_{\rm bm} = \mathbb{P}(T > \tau),
\end{align*}
where $T$ is an exponential random variable with mean $\frac{1}{v\mu_{\rm s,b}}$, with $T$ interpreted as the time between two consecutive SSBs. 
Note that $\frac{1}{v\mu_{\rm s,b}}$ is the average time between two consecutive beam reselections. Hence, the average time between two SSBs is equal to the average time between two consecutive beam reselections. Based on the probability of the beam misalignment, the antenna gain $G_{0}$ of the serving BS is given as
\begin{align}
G_{0} = \begin{cases}
               G_{{\rm m},  n} & \mathrm{w.p.}~1-p_{\rm bm}\\
               G_{{\rm s},  n}  & \mathrm{w.p.}~p_{\rm bm}.
            \end{cases}
            \label{eq:gain_serv}
\end{align}

\subsection{Blockage and Channel Model}

The presence of obstacles leads to LOS and NLOS propagation between the typical MT and a BS. We adopt a LOS ball model~\cite{Bai} to capture the effect of blockages. In particular, the propagation between a BS and the typical MT separated by distance $d$ is LOS if $d < R_{\rm c}$
where $R_{\rm c}$ is the maximum distance for LOS propagation. 
The LOS and NLOS channel conditions induced by the blockage effect are characterized by different path loss exponents, denoted by $\alpha_{\rm L}$ and $\alpha_{\rm N}$, respectively. Typical values of these path loss exponents are $\alpha_{\rm L} \in [1.8, 2.5]$ and $\alpha_{\rm N} \in [2.5, 4.7]$.

The channel follows Rayleigh fading with unit mean power gain. Let $h_{X}$ denote the channel power gain from the BS at location $X \in \Phi$ to the MT.
Note that, the random variables $h_{X}$ are i.i.d. exponential with unit mean, i.e., $h_{X} \sim \exp(1)$. Let $|X|$ denote the distance between a BS at $X \in \Phi$ and the typical MT located at the origin. We consider the standard power-law path loss model with path loss function as 
\begin{align}
l(X) = \begin{cases}
               K|X|^{-\alpha_{\rm L}} & \mathrm{if}~|X| < R_{\rm c}\\
               K|X|^{-\alpha_{\rm N}} & \mathrm{if}~|X| \geq R_{\rm c},
            \end{cases}
\end{align}
where $K = \left(\frac{c}{4\pi f_{\rm c}}\right)^2$ is a frequency-dependent constant with $c$ being the speed of light and $f_{\rm c}$ the carrier frequency. We assume that the typical MT can receive from its serving BS (also the nearest BS) in both LOS and NLOS conditions.

\subsection{Signal-to-Interference-Noise Ratio (SINR)}

When the typical MT is associated with the BS located at $X_0 \in \Phi$ with antenna gain $G_0$ as in \eqref{eq:gain_serv}, the SINR at the typical MT is given by
\begin{align}
\mathsf{SINR}_{n} = \frac{PG_{0}h_{X_0} l(|X_0|)}{\sigma^2 + I_{n} }, 
\label{eq:sir}
\end{align}
where $P$ is the transmit power of a BS.  Also, $\sigma^2 = WN_0$ is the noise power, where $W$ and $N_0$ are the bandwidth and the noise spectral density, respectively. In the denominator of \eqref{eq:sir},  $I_{n}$ is the interference power at the typical MT given by
\begin{align}
I_{n} = \sum_{X \in \Phi\setminus \lbrace X_0\rbrace} P G_{n}h_{X}l(|X|).
\label{eq:intf}
\end{align}
Since the beams of all BSs are oriented towards their respective MTs, the direction of arrivals between interfering BSs and the typical MT is distributed uniformly in $[-\pi, \pi]$. Thus, the gain $G_n$ of an interfering BS is equal to $G_{{\rm m}, n}$ with probability $\varphi_n/2\pi$ and $G_{{\rm s}, n}$ with probability $1- \varphi_n/2\pi$, where $\varphi_n$ is the beamwidth given by \eqref{eq:bw}.

\section{Ergodic Shannon Rate}
We are interested in calculating the downlink ergodic Shannon rate at the typical MT, which is given as
\begin{align}
\mathcal{R}_{n} = W\mathbb{E}\left[\log\left(1 + \mathsf{SINR}_{n}\right)\right],
\end{align}
where $\mathbb{E}(\cdot)$ denotes expectation.


\begin{definition}[Success Probability]
The success probability $p_{\rm s}$ of the typical MT is the probability that the SINR at the typical MT exceeds a predefined threshold. Mathematically,
\begin{align}
p_{\rm s}(n,\beta) \triangleq \mathbb{P}(\mathsf{SINR}_{n} > \beta),
\end{align}
where $\beta > 0$ is the predefined SINR threshold, which also parametrizes the transmission rate.
\end{definition}

\begin{lemma}
\label{lem:suc_prob}
Let
{\small
\begin{align*}
F(\alpha_{\rm S}, \alpha_{\rm I}, w) \triangleq 2\pi \lambda \hspace*{-1mm}\left(1- \frac{p_{I,n}}{1+\frac{\beta r^{\alpha_{\rm S}} G_{{\rm m}, n} }{G_0 w^{\alpha_{\rm I}}}} -  \frac{1-p_{I,n}}{1+\frac{\beta r^{\alpha_{\rm S}} G_{{\rm s}, n} }{G_0 w^{\alpha_{\rm I}}}}\hspace{-1mm}\right)\hspace{-1mm}w,
\end{align*}}
where $p_{I,n} = \varphi_n/2\pi$. The success probability $p_{\rm s}(n,\beta)$ is 
\begin{align}
p_{\rm s}(n,\beta) = (1- p_{\rm bm}) q_{\rm s}(n,\beta, G_{{\rm m},n})+ p_{\rm bm}q_{\rm s}(n,\beta, G_{{\rm s},n}),
\end{align}
where $p_{\rm bm}$ is given by \eqref{eq:p_bm} and\vspace{-3mm}

{\small
\begin{align}
&q_{\rm s}(n,\beta, G_0) = \int_{0}^{R_{\rm c}}f_{R}(r) \exp\left(-\frac{\beta r^{\alpha_{\rm L}}\sigma^2}{PKG_0}\right) \nonumber \\
&\times \exp\left(\hspace*{-1mm}-\int_{r}^{R_{\rm c}}\hspace*{-1mm}F(\alpha_{\rm L}, \alpha_{\rm L}, w) \mathrm{d}w -\hspace*{-1mm} \int_{R_{\rm c}}^{\infty}\hspace*{-1mm}F(\alpha_{\rm L}, \alpha_{\rm N}, w)\mathrm{d}w\right) \mathrm{d}r \nonumber \\
&+ \int_{R_{\rm c}}^{\infty}\exp\left(-\frac{\beta r^{\alpha_{\rm N}}\sigma^2}{PKG_0} - \int_{r}^{\infty} F(\alpha_{\rm N}, \alpha_{\rm N}, w)\mathrm{d}w\right)f_{R}(r)\mathrm{d}r
\label{eq:suc_prob}
\end{align}}
\vspace*{-3mm}

\noindent with $f_R(r) = 2\pi \lambda r e^{-\lambda \pi r^2}$ where $R$ is the distance to the nearest BS.
\end{lemma}

\begin{IEEEproof}
The proof is given in Appendix~\ref{app:suc_prob}.
\end{IEEEproof}

\begin{theorem}
Imposing the limit $Q_{\rm max}$ on the maximum achievable SINR stemming from RF imperfections and modulation schemes, the ergodic Shannon rate per unit time is
\begin{align}
\mathcal{R}_{n}  = W\int_{0}^{Q_{\rm max}} \frac{p_{\rm s}(n,z)}{z+1}~\mathrm{d}z,
\label{eq:av_rate}
\end{align}
where $p_{\rm s}(n,z)$ is given by \eqref{eq:suc_prob}.
\end{theorem}
\begin{IEEEproof}
The proof follows directly from
\begin{align}
\mathcal{R}_{n} &= W\mathbb{E}[\log(1+\min(\mathsf{SINR}_{n}, Q_{\rm max}))]. \nonumber
\end{align}
\end{IEEEproof}

\section{Beam (Re)Selection}

\subsection{Beam Selection during BS handovers}
When the MT performs a BS handover, it selects a beam with the new BS. We know from \cite{FB_cell_crossings} that, for the Poisson-Voronoi tessellation, the linear intensity of \textit{cell} boundary crossing, i.e., BS handover, is $\mu_{\rm s,c} = \frac{4\sqrt{\lambda}}{\pi}$. Hence, the time intensity of BS handover (or, equivalently, beam selection) is
\begin{align}
\mu_{\rm c} = \frac{4\sqrt{\lambda}}{\pi}v.
\end{align}

\subsection{Beam Reselection}
\label{sec:beam_resel}

The beam reselection occurs at a beam boundary within the Voronoi cell of a BS. In Fig.~\ref{fig:PVC}, the locations of beam reselections are denoted by brown filled circles. 
We are interested in calculating the average number of beam reselections the typical MT performs per unit length. We also calculate the time intensity of beam reselection, i.e., the average rate of beam reselections.


\begin{theorem}[Intensity of beam reselection]
For $2^n$ beams and PPP of intensity $\lambda$, the linear intensity $\mu_{\rm s,b}$ of beam reselection for the typical MT moving on a straight line with speed $v$ is $\frac{2^n\sqrt{\lambda}}{\pi}$. The time intensity $\mu_{\rm t,b}$ of beam reselection is $\mu_{\rm t,b}(n) = \frac{2^n\sqrt{\lambda}}{\pi}v$.
\label{thm:int_beam}
\end{theorem}
\begin{IEEEproof}
The proof is given in Appendix~\ref{app:int_beam}.
\end{IEEEproof}

\begin{remark}
When taken into account the average number of beam reselections that are skipped between two consecutive SSBs, the effective time intensity of beam reselection becomes
\begin{align}
\mu_{\rm b}(n) =  \frac{1}{\max\left(\tau, \frac{1}{\mu_{\rm t, b}(n)}\right)},
\end{align} 
where $\tau$ is the SSB periodicity.
\end{remark}

\section{Time Overhead due to Beam (Re)Selection}

BS handovers and beam reselections may result in significant overheads in terms of time as a consequence of the time spent in beam sweeping and alignment, respectively. Such an overhead reduces the time available for data transmissions, in turn reducing the ergodic Shannon rate.

The typical MT moves on a straight line across different beams within the Voronoi cell of a BS as well as across Voronoi cells of different BSs. Hence, the two components that contribute to the time overhead are:

%

\begin{enumerate}
\item The time $T_{\mathrm{c}}$ for beam sweeping after each BS handover, i.e., cell boundary crossing, which includes the periodic SSB measurement, receiver processing time for the SSBs, and the handover interruption time due to cell switch and radio resource control (RRC) reconfiguration.
\item The time $T_{\mathrm{b}}$ for beam alignment after each beam reselection within the Voronoi cell of a BS, which includes the periodic SSB measurement and the receiver processing time for the SSBs.

\end{enumerate}
Thus, the total average overhead per unit time is 
\begin{align}
T_{\mathrm{o}}(n) =  \mu_{\rm b}(n)T_{\mathrm{b}} +\mu_{\rm c} T_{\mathrm{c}}.
\end{align}
The effective area spectral efficiency (ASE) per unit time is
\begin{align}
\mathcal{R}_{\mathrm{eff}}(n) = \lambda \mathcal{R}_{n}(1-T_{\rm o}(n))^{+},
\end{align}
where $\mathcal{R}_{n}$ is given by \eqref{eq:av_rate} and $(A)^{+} = \max(0, A)$.

Our objective is to find the integer $n$ that maximizes the effective ASE per unit time, i.e.,
\begin{align}
n_{*} = \argmax_{n \in \mathbb{N}} ~\mathcal{R}_{\mathrm{eff}}(n).
\end{align}
The value of $n_{*}$ can easily be found by a linear search.

\begin{table}[]
\caption{Values of network parameters~\cite{spec}}\vspace*{-1mm}
\begin{tabular}{|l|l|l|}
\hline
  \hspace{-1mm}Parameter              &  \hspace{-1mm}FR1                    &  \hspace{-1mm}FR2                                            \\\hline
 \hspace{-1mm}Carrier frequency ($f_{\rm c}$)    &  \hspace{-1mm}$3.5~\mathrm{GHz}$                &  \hspace{-1mm}$28 ~\mathrm{GHz}                                        $ \\\hline
 \hspace{-1mm}Bandwidth ($W$)              &  \hspace{-1mm}$100 ~\mathrm{MHz}$                &  \hspace{-1mm}$400~\mathrm{MHz}$                                        \\\hline
 \hspace{-1mm}Noise density ($N_0$)           &  \hspace{-1mm}$-174 ~\mathrm{dBm/Hz}$            &  \hspace{-1mm}$-174 ~\mathrm{dBm/Hz}                                   $ \\\hline
 \hspace{-1mm}Transmit power ($P$)             &  \hspace{-1mm}$43~ \mathrm{dBm}$                 &  \hspace{-1mm}$36~\mathrm{dBm}                                        $ \\\hline
 \hspace{-1mm}Beam reselection overhead ($T_{\rm b}$) \hspace{-3mm}           &  \hspace{-1mm}$23~\mathrm{ms}$                   &  \hspace{-1mm}$23~\mathrm{ms}$                                           \\\hline
 \hspace{-1mm}Cell handover overhead ($T_{\rm c}$) \hspace{-3mm}          &  \hspace{-1mm}$43~\mathrm{ms} $                 &  \hspace{-1mm}$43~\mathrm{ms}$                                          \\\hline
 \hspace{-1mm}SSB periodicity ($\tau$)          &  \hspace{-1mm}$20~\mathrm{ms}$                  &  \hspace{-1mm}$20 ~\mathrm{ms                                        } $ \\\hline
 \hspace{-1mm}MT speed ($v$)            &  \hspace{-1mm}$[3, 30, 120]~\mathrm{km/h}$ \hspace*{-2mm} &  \hspace{-1mm}$[3, 30]~\mathrm{km/h}                              $ \\\hline
 \hspace{-1mm}Inter-site distance (ISD)            &  \hspace{-1mm}$[250, 500, 1000] ~\mathrm{m}$ \hspace{-9mm} &  \hspace{-1mm}$[75, 125, 250{]} ~\mathrm{m}$                           \\\hline

 \hspace{-1mm}Maximum SINR ($Q_{\rm max}$)            &  \hspace{-1mm}$30 ~\mathrm{dB}$ &  \hspace{-1mm}$30 ~\mathrm{dB} $                          \\\hline
 \hspace{-1mm}Path loss exponent ($\alpha$)       &  \hspace{-1mm}3.5                    &  \hspace{-1mm}$\alpha_{\rm L} = 1.9$, $\alpha_{\rm N} = 3.5$ \hspace{-3mm} \\\hline
 \hspace{-1mm}Blockage model &  \hspace{-1mm}Implicit (NLOS)                   &  \hspace{-1mm}LOS ball                                      \\\hline
  \hspace{-1mm}LOS ball radius ($R_{\rm c}$)            &  \hspace{-1mm}$-$&  \hspace{-1mm}$75 ~\mathrm{m}                          $ \\\hline
\end{tabular}
\label{tab:1}\vspace*{-3mm}
\end{table}

\section{Numerical Results and Discussions}
To validate our proposed model, we consider a $5$G NR-compliant radio access network operating in a dense urban macro/pico cell scenario. A summary of the model parameters for FR1 (sub-$6 ~\mathrm{ GHz}$) and FR2 (above $6 ~\mathrm{GHz}$) network deployments is provided in Table~\ref{tab:1}.

To explore the best potential of $5$G NR networks in operational bands FR1 and FR2, we use the maximum bandwidth allowed as per $5$G NR Release 15\cite{3gpp.38.101-1, 3gpp.38.101-2}, namely $100 ~\mathrm{MHz}$ for FR1 and $400~\mathrm{MHz}$ for FR2. 
As the inter-site (or, equivalently inter-base station) distances (ISDs) for FR1 are expected to be larger due to lower frequency-dependent path loss, we choose a transmission power $P$ = $43 ~\mathrm{dBm}$. For FR2, we can expect smaller ISDs due to (a) higher attenuation loss at higher carrier frequencies, and (b) because massive MIMO and high beamforming gains imposes challenges in terms of RF exposure and EMF limitations. 
Thus we decrease $P$ to $36 ~\mathrm{dBm}$. 

One important network planning decision is the number of beams per BS. In our proposed model, this is related to the choice of the values for $n$. The main and side lobe gains depend on the number $2^n$ of beams. In particular, when increasing the number of beams, the main lobe gain increases while the side lobe gain decreases. Below, we assume the following antenna gains: the main lobe gain is $G_{{\rm m}, n} = 2^n$, while the side lobe gain is $G_{{\rm s}, n} = \frac{1}{2^n}$.

Another important parameter is the intensity $\lambda$ of BSs, which is directly related to the average cell size and the ISD. For an intensity $\lambda$ of BSs, the average cell size is $1/\lambda$~\cite{FB_book}. The average cell radius in a network is defined as the radius $r_{\rm cell}$ of a ball having the same average area as the cell: $1/\lambda$. Then, the average ISD is $2r_{\rm cell} = 2/(\sqrt{\pi \lambda})$. Hence we simply adjust the value of the intensity $\lambda$ of BSs to represent different ISD scenarios. For FR1 bands, we analyze the following ISDs: $1000 ~\mathrm{m}$, $500 ~\mathrm{m}$, and $250~\mathrm{m}$. As for FR2 bands, the effect of propagation attenuation requires us to decrease the ISD. Hence we consider the following ISDs: $250 ~\mathrm{m}$, $125 ~\mathrm{m}$, and $75 ~\mathrm{m}$. 


\begin{figure}
\centering
\includegraphics[scale=.45]{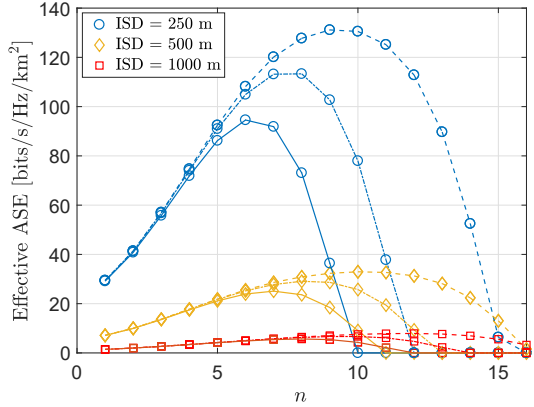}\vspace*{-1mm}
\caption{\small \textbf{FR1}: Effect of average inter-site distance (ISD) on the effective ASE for different $n$. Dashed line: $v = 3~\rm{km/h}$,
Dashed-dotted line: $v = 30~ \rm{km/h}$, and
Solid line: $v = 120~ \rm{km/h}$.}
\label{fig:diff_ISD_diff_vel_FR1}\vspace*{-3mm}
\end{figure}

\begin{figure}
\centering
\includegraphics[scale=.45]{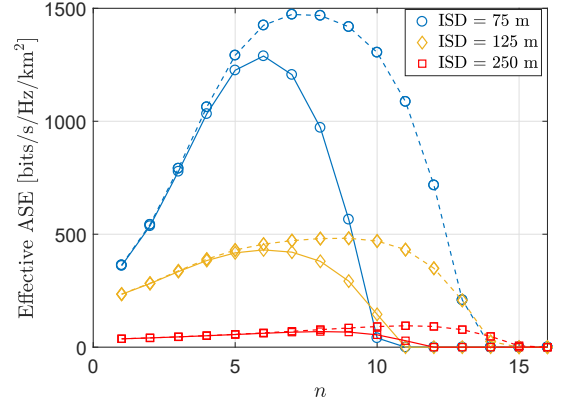}\vspace*{-1mm}
\caption{\small \textbf{FR2}: Effect of average inter-site distance (ISD) on the effective ASE for different $n$. Dashed line: $v = 3~\rm{km/h}$ and 
Solid line: $v = 30~ \rm{km/h}$.}
\label{fig:diff_ISD_diff_vel_FR2}\vspace*{-3mm}
\end{figure}

\begin{figure}
\centering
\includegraphics[scale=.45]{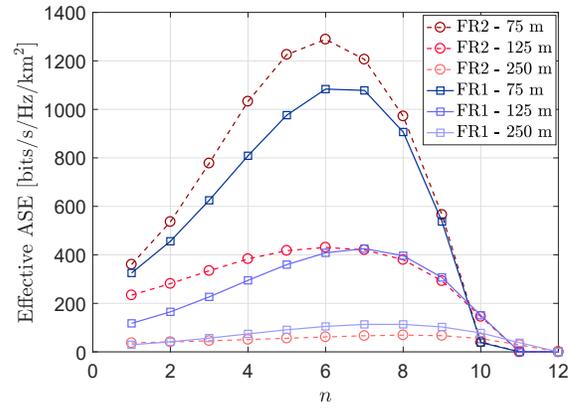}
\caption{\small Comparison between FR1 and FR2 deployments for different ISDs with $v  = 30~\rm{km/h}$.}
\label{fig:comp_FR1_FR2}\vspace*{-2mm}
\end{figure}

In Figs.~\ref{fig:diff_ISD_diff_vel_FR1},  \ref{fig:diff_ISD_diff_vel_FR2}, and \ref{fig:comp_FR1_FR2}, we evaluate the performances of FR1 and FR2 deployments, for different MT speeds $v$ and ISDs. The following discussions are valid for Figs. \ref{fig:diff_ISD_diff_vel_FR1}, \ref{fig:diff_ISD_diff_vel_FR2}, and \ref{fig:comp_FR1_FR2}. 

In both FR1 and FR2 deployments, for a given ISD, as $n$ (and hence the number $2^n$ of beams) increases, the beamforming gain increases and the interference from interfering BSs decreases due to narrower beams. But, at the same time, for a given $v$, the typical MT performs more frequent beam reselections due to a smaller beamwidth, increasing the time overhead. The negative impact of the increased time overhead gradually dominates the beamforming gain. Also, with an increase in $n$, the probability of beam misalignment (given by \eqref{eq:p_bm}) increases as the MT is more likely to move into the beamwidth of another beam between two consecutive SSB measurements. As a result of this tussle,
 we observe that there is a value $n_{*}$ of $n$ that maximizes the effective ASE.

Similarly, with an increase in $v$, the MT crosses beam and cell boundaries more frequently resulting in a higher number of beam reselections and BS handovers, respectively. Also, the probability of beam misalignment increases. Hence, with an increase in $v$, the optimal value $n_{*}$ decreases, and for a fixed $n$, the effective ASE decreases. 

Note that a smaller ISD means a denser cellular network. As the ISD increases, the average cell size increases. Thus, BSs need to increase the number $2^n$ of beams for a higher beamforming gain (equivalently, a smaller beamwidth) and smaller interference. Hence, the value of $n_{*}$ increases with ISD. For a smaller $n$, 
the time overhead associated with beam reselections is small. Hence, a smaller ISD results in a higher effective ASE due to a smaller distance between the typical MT and its serving BS. This dominates the negative effects of increased interference power due to a denser deployment of BSs and increased time overhead due to more frequent BS handovers. But, for a large $n$, both beam reselections and BS handovers happen more frequently for a smaller ISD, and the network with a larger ISD achieves a higher effective ASE.

When comparing beamwidths for FR1 with those for FR2 given the same $n$ value, we can expect that, even though the angular beamwidth for a given $n$ is the same, the linear width of the beams will be larger in FR1 than that in FR2 due to a \textit{usually} larger cell radius in the former.\footnote{Although nothing prohibits small ISDs for FR1 (e.g., Wi-Fi operation in $2.4~\mathrm{ GHz}$ and $5~\mathrm{ GHz}$), pico/femto cell deployment models are the best fit for mmWave deployments.} Thus, it is expected that MTs operating in FR1 are less penalized due to overheads associated with beam reselections and BS handovers. Also, due to a larger ISD, the \textit{distance-based} path loss when operating in FR1 is higher than when operating in FR2. Hence for the same value of $n$, we can expect that the interference in FR1 is smaller than that in FR2 despite a bit higher transmit power in FR1. This indicates that the additional received interference from a higher number of beams will not be as intense in FR2 as in FR1. As shown in Fig.~\ref{fig:comp_FR1_FR2}, this behavior is captured by our model, where the optimal value $n_{*}$ for the ISD $= 250 ~\mathrm{m}$ in FR1 ($n_{*} = 8$) is higher than that in FR2 ($n_{*} = 6$) with ISD $=75~\rm{m}$ or in FR2 with ISD $=125~\rm{m}$ ($n_{*} = 7$).

Also, as Fig.~\ref{fig:comp_FR1_FR2} shows, when we compare the performances of FR1 and FR2 deployments for the same ISD, we observe an interesting trade-off. Smaller ISDs (e.g., $75~\rm{m}$) benefit FR2 irrespective of the value of $n$ as, for a given critical LOS distance $R_{\rm c}$, it is more likely that the serving BS has LOS propagation to the MT (so smaller path loss) boosting signal power. Recall that, in FR1, even the serving BS always has NLOS propagation to the MT. As a result, the impact of increased time overheads with $n$ on the effective ASE is less in FR2 than that in FR1. As the ISD increases, the probability of serving BS lying outside the LOS ball increases, in turn, reducing signal power significantly in FR2. Thus, when combined with higher path loss at higher frequencies in FR2, for higher values of $n$, the impact of time overheads is relatively higher in FR2 than FR1. Note that, as a result of the tussle between competing effects, the values of $n_{*}$ in FR1 and FR2 are close to each other for the same ISD.

Overall, the numerical results  discussed in this section validate the claim that the proposed model fits as a tool for system-level evaluation of network planning decisions in beam-based access networks such as $5$G NR.

\section{Conclusions and Future Directions}

We presented a system-level stochastic geometry model for a $5$G NR radio access network that allows one to capture essential scenario characteristics (e.g., operational frequency, blockage characteristics), technological features (e.g., beamforming configuration, delay overheads), and network deployment choices (e.g., ISD, carrier bandwidth). We analyzed the application of the model for a dense urban macro/pico cell scenario for FR1 and FR2. We demonstrated that the model accurately captures the existing trade-offs between ISDs, path loss, interference, and signaling overhead due to beam management. This also shows that the stochastic geometry model can be effective for conducting system-level analysis of beam-based radio access networks such as $5$G NR.

In the future, we will extend our model to capture system-level impact of multiple MT panels and/or multiple MT beams. We also aim to improve the model to incorporate sectorized BSs with antenna panels and their effects on beam shaping/gains, MT blockage models, and others. Finally, we intend to extend our model to capture other trade-offs involved in designing a beam-based radio access network. For instance, with an increase in the number of beams, the average beam time-of-stay (ToS), i.e., the time a user remains with a given beam before switching to another beam, decreases, which impacts the available time for performing channel estimation, link adaptation, and power control.


\appendices

\section{Proof of Lemma~\ref{lem:suc_prob}}
\label{app:suc_prob}

For the given antenna gain $G_0$ at the serving BS, the success probability is
\begin{align}
q_{\rm s}(n, \beta, G_0) & \triangleq \mathbb{P}(\mathsf{SINR}_n > \beta \mid G_0).
\label{eq:ps_app}
\end{align}
Depending on whether the serving BS lies within the LOS ball of radius $R_{\rm c}$ or not, we can write \eqref{eq:ps_app} as
\begin{align}
q_{\rm s}(n,\beta, G_0) &= \int_{0}^{R_{\rm c}} \mathbb{P}(\mathsf{SINR}_n > \beta \mid r, G_0)f_{R}(r)\mathrm{d}r \nonumber\\
&+ \int_{R_{\rm c}}^{\infty} \mathbb{P}(\mathsf{SINR}_n > \beta \mid r, G_0)f_{R}(r)\mathrm{d}r,
\end{align}
where $f_{R}(r) = 2\pi \lambda r \exp(-\lambda \pi r^2)$ is the probability density function of the distance $R$ of the typical MT to the nearest BS. When $0 < r < R_{\rm c}$, we have
\begin{align}
\mathbb{P}(\mathsf{SINR}_n > \beta \mid r, G_0) &= \mathbb{P}\left(
\frac{PKG_0 h_{X_0}r^{-\alpha_{\rm L}}}{N_0 + I_n} > \beta \mid r\right)\nonumber\\
& = \mathbb{P}\left(h_{X_0} > \frac{\beta r^{\alpha_{\rm L}} (N_0 + I_n)}{PKG_0} \mid r\right) \nonumber \\
& = e^{-\frac{\beta r^{\alpha_{\rm L}}N_0}{PKG_0}}\mathbb{E}\left[\exp\left(-\frac{\beta r^{\alpha_{\rm L}} I_n}{PKG_0}\right)\right].
\end{align}
We have
\begin{align}
&\mathbb{E}\left[\exp\left(-\frac{\beta r^{\alpha_{\rm L}} I_n}{PKG_0}\right)\right] \nonumber \\
& = \mathbb{E}\left[\exp\left(-\frac{\beta r^{\alpha_{\rm L}} \sum_{X \in \Phi\setminus \lbrace X_0\rbrace} P G_{n} h_{X}l(|X|)}{PKG_0}\right)\right] \nonumber \\
&\overset{(\mathrm{a})}{=} \mathbb{E}\left[\prod_{X \in \Phi\setminus \lbrace X_0\rbrace} \mathbb{E}\left(\exp\left(-\frac{\beta r^{\alpha_{\rm L}} G_{n} h_{X}l(|X|)}{KG_0}\right)\right)\right] \nonumber \\
&\overset{(\mathrm{b})}{=}\mathbb{E}\left[\prod_{X \in \Phi\setminus \lbrace X_0\rbrace} \mathbb{E}\left(\frac{1}{1 + \frac{\beta r^{\alpha_{\rm L}}l(|X|)G_n}{KG_0}}\right)\right]\nonumber \\
& \overset{(\mathrm{c})}{=} \mathbb{E}\left[\prod_{X \in \Phi\setminus \lbrace X_0\rbrace}\hspace*{-2mm} \left(\frac{p_{I,n}}{1 + \frac{\beta r^{\alpha_{\rm L}}l(|X|)G_{\rm m, n}}{KG_0}} + \frac{1-p_{I,n}}{1 + \frac{\beta r^{\alpha_{\rm L}}l(|X|)G_{\rm s,n}}{KG_0}}\hspace*{-1mm}\right)\hspace*{-1mm}\right]\hspace*{-1mm}, \nonumber
\end{align}
where $(\mathrm{a})$ follows from the i.i.d. nature of fading random variables, $(\mathrm{b})$ follows from averaging over the channel power gains on interfering channels, and $(\mathrm{c})$ follows from the fact that the interference from an interfering BS at $X\in \Phi$ comes from its main lobe with probability $p_{I,n}$ and from its side lobe with probability $1-p_{I,n}$ with $p_{I,n} = \varphi_n/2\pi$ the probability that the typical MT lies within the beamwidth of the main lobe of an interfering BS.

From the probability generating functional (PGFL) of PPP, it follows that

{\small\begin{align}
&\mathbb{E}\left[\exp\left(-\frac{\beta r^{\alpha_{\rm L}} I_n}{PKG_0}\right)\right] \nonumber \\
&= \exp\hspace*{-1mm}\left(\hspace*{-1mm}-2\pi \lambda \hspace*{-1mm}\int_{r}^{\infty} \hspace*{-2mm}\left(\hspace*{-1mm} 1- \frac{p_{I,n}}{1+\frac{\beta r^{\alpha_{\rm L}} l(w)G_{{\rm m}, n} }{KG_0 }} -  \frac{1-p_{I,n}}{1+\frac{\beta r^{\alpha_{\rm L}} l(w)G_{{\rm s}, n} }{KG_0}}\hspace{-1mm}\right)\hspace{-1mm}w \mathrm{d}w\hspace*{-1mm}\right) \nonumber \\
&= \exp\left(\hspace*{-1mm}-\int_{r}^{R_{\rm c}}\hspace*{-1mm}F(\alpha_{\rm L}, \alpha_{\rm L}, w) \mathrm{d}w -\hspace*{-1mm} \int_{R_{\rm c}}^{\infty}\hspace*{-1mm}F(\alpha_{\rm L}, \alpha_{\rm N}, w)\mathrm{d}w\right),
\label{eq:main_side}
\end{align}}
where
{\small
\begin{align*}
F(\alpha_{\rm S}, \alpha_{\rm I}, w) \triangleq 2\pi \lambda \hspace*{-1mm}\left(1- \frac{p_{I,n}}{1+\frac{\beta r^{\alpha_{\rm S}} G_{{\rm m}, n} }{G_0 w^{\alpha_{\rm I}}}} -  \frac{1-p_{I,n}}{1+\frac{\beta r^{\alpha_{\rm S}} G_{{\rm s}, n} }{G_0 w^{\alpha_{\rm I}}}}\hspace{-1mm}\right)\hspace{-1mm}w.
\end{align*}}

Similarly, we can obtain an expression for $\mathbb{P}(\mathsf{SINR}_n > \beta \mid r)$ when $R_{\rm c} \leq r < \infty$. 

\begin{figure}
\centering
\includegraphics[scale=.33]{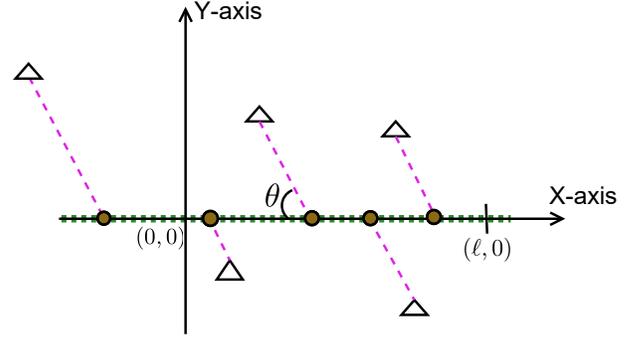}
\caption{At most one beam reselection for a BS. $\bigtriangleup$ : Base station (BS), brown filled circle : Beam reselection location, dashed lines : Beam boundaries, and dotted line : Path of MT along the $X$-axis.}
\label{fig:BM_model}
\end{figure} 

\section{Proof of Theorem~\ref{thm:int_beam}}
\label{app:int_beam}
The beam reselections occur at beam boundaries of a BS. In Fig.~\ref{fig:PVC}, the locations of beam reselections are denoted by brown filled circles. As shown in Fig.~\ref{fig:BM_model}, let $\theta$ denote the angle of a beam boundary with respect to the direction of the motion of the MT. This angle $\theta$ is distributed uniformly at random in $[0,\pi]$. In Fig.~\ref{fig:BM_model}, a `brown filled circle' denotes a point along the $X$-axis where the typical MT reselects the beam. Let $\Psi \subset \mathbb{R}$ denote the point process of beam reselection events, where the points of $\Psi$ are indicated by `brown filled circles.' We are interested in calculating the average number of beam reselections the typical MT performs per unit length, which is equivalent to calculating the intensity of $\Psi$.

Without loss of generality, let us consider the motion of the typical MT in the interval $[0,\ell]$, which corresponds to the motion of the typical MT from $(0, 0)$ to $(\ell, 0)$ along the $X$-axis. First, we calculate the average number of beam reselections when there is at most one beam reselection corresponding to a BS. This happens when there are two beams of the same size for a BS. Using the result for two beams, we can obtain the intensity of beam reselections corresponding $2^n$ beams of a BS.

\begin{figure}
\centering
\includegraphics[scale=.27]{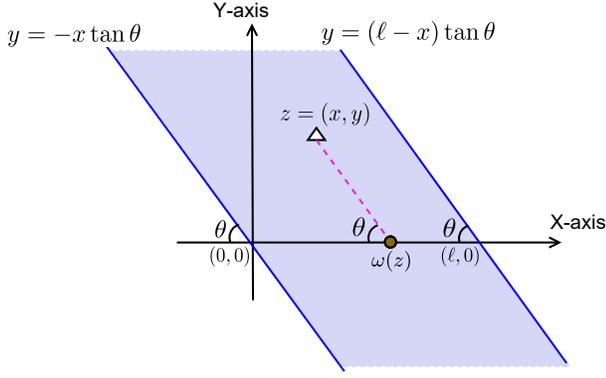}
\caption{Strip $\mathcal{S}$ (shaded area) between two blue solid lines with angle $\theta$ with the $X$-axis. $\bigtriangleup$ : BS, brown filled circle : Beam reselection location, and dashed line : Beam boundary.}
\label{fig:strip}
\end{figure}

During the interval $[0,\ell]$, the typical MT may move through the Voronoi cells of multiple BSs. As shown in Fig.~\ref{fig:strip}, let $\omega(X)$ denote the point of the beam reselection corresponding to the BS located at $X \in \Phi$. The event of beam reselection corresponding to a BS at $X\in \Phi$ occurs when the following two events occur simultaneously:
\begin{enumerate}

\item the point of the beam reselection lies in the Voronoi cell of the BS at $X\in \Phi$, i.e., $\omega(X) \in V_{\Phi}(X)$ where $V_{\Phi}(X)$ is the Voronoi cell of the BS located at $X \in \Phi$.
\item the point of the beam reselection lies on the line connecting $(0, 0)$ and $(\ell,0)$, i.e, $\omega(X) \in [0,\ell]$. 
\end{enumerate}
Consequently, conditioning on $\theta$, the average of number of beam reselections in $[0,\ell]$ is
\begin{align*}
\mathbb{E}(\Psi[0,\ell]\mid \theta)  = \mathbb{E}\left(\displaystyle\sum_{X \in \Phi} \boldsymbol{1}_{\omega(X) \in V_{\Phi}(X)}\boldsymbol{1}_{\omega(X) \in [0,\ell]}\right),
\end{align*}
where $\boldsymbol{1}$ is the indicator function. Conditioning on the fact that there is a BS at location $z \in \mathbb{R}^2$ and using the Campbell's theorem~\cite{FB_book}, it follows that
\begin{align}
\mathbb{E}(\Psi[0, \ell]\mid \theta)  =\lambda \int_{\mathbb{R}^2}\mathbb{P}_{z}(\omega(z) \in V_{\Phi}(z))~\boldsymbol{1}_{\omega(z) \in [0, \ell]}~\mathrm{d}z, 
\label{eq:campbell}
\end{align}
where $\mathbb{P}_z(\cdot)$ denotes the Palm probability.

\textbf{Event $\boldsymbol{\omega(z) \in V_{\Phi}(z)}$}: This event occurs when there is no BS closer to the typical MT than the one at location $z$. This is equivalent to the event that there is no BS in the ball of radius $\|z - \omega(z)\|$ centered at $\omega(z)$. Since the BS point process is a PPP with intensity $\lambda$, we have
\begin{align}
\mathbb{P}_{z}(\omega(z) \in V_{\Phi}(z)) &= \exp\left(-\lambda \pi \|z - \omega(z)\|^2\right) \nonumber \\
& = \exp\left(-\frac{\lambda \pi y^2}{\sin^2 \theta}\right),
\label{eq:voronoi_event}
\end{align}
where $\|z - \omega(z)\| = \frac{y}{\sin \theta}$ since $z  = (x, y)$. 

\textbf{Event $\boldsymbol{\omega(z) \in [0,\ell]}$}: This event occurs if the BS at $z$ lies within the strip $\mathcal{S}$ between two lines passing through the origin $(0,0)$ and $(\ell,0)$ at angle $\theta$ as shown in Fig.~\ref{fig:strip}. Equivalently, by representing $z$ in Euclidean coordinates as $z = (x,y)$, it follows that 
\begin{align}
\boldsymbol{1}_{\omega(z) \in [0,\ell]}~\mathrm{d}z =\boldsymbol{1}_{z \in \mathcal{S}} ~\mathrm{d}z =  \boldsymbol{1}_{(x,y) \in \mathcal{S}}~\mathrm{d}x\mathrm{d}y.
\label{eq:line_event}
\end{align}
The left line passing through the origin at an angle $\theta$ can be given as $y = -x\tan\theta$, while the right line passing through $(\ell,0)$ at an angle $\theta$ can be given as $y = (\ell-x)\tan \theta$. Thus, the BS at $z$ lies in the strip $\mathcal{S}$ if 
\begin{align}
-\infty < y < \infty ~~~\text{and}~~~ -\frac{y}{\tan \theta} \leq x \leq \ell-\frac{y}{\tan \theta}.
\label{eq:boundary_strip}
\end{align}

From \eqref{eq:voronoi_event}, \eqref{eq:line_event}, and \eqref{eq:boundary_strip}, we can express \eqref{eq:campbell} as
\begin{align}
\mathbb{E}(\Psi[0,\ell]\mid \theta)  &= \hspace*{-1mm}\lambda \int_{y = -\infty}^{\infty}\hspace*{-4mm}\mathrm{d}y\int_{x = -\frac{y}{\tan \theta}}^{\ell -\frac{y}{\tan \theta}}~\hspace*{-3mm}\exp\left(-\frac{\lambda \pi y^2}{\sin^2 \theta}\right)\mathrm{d}x \nonumber \\
& =\ell \left(\sqrt{\lambda} |\sin \theta|\right) .
\end{align}

Since $\theta$ is uniformly distributed in $[0,\pi]$, averaging over $\theta$ yields
\begin{align}
\mathbb{E}(\Psi[0,\ell]) 
& = \left(\frac{2\sqrt{\lambda}}{\pi}\right)\ell.
\end{align}
Hence, for the case of two beams, the linear intensity of beam reselections is $\frac{2\sqrt{\lambda}}{\pi}$.

For $2^n$ beams with $n \in \mathbb{N}$, there are $2^{n-1}$ possibilities of beam reselections corresponding to the serving BS. In this case, we can obtain the average number of beam reselections by superimposing the beam reselection events corresponding to each beam boundary crossing considered earlier in this proof. Hence, the linear intensity of beam reselections for $2^n$ beams is
\begin{align}
\mu_{\rm s,b}(n) = 2^{n-1}\frac{2\sqrt{\lambda}}{\pi} =\frac{2^n\sqrt{\lambda}}{\pi}.
\label{eq:int_beam}
\end{align}
By considering the speed $v$ of the typical MT, we get the time intensity $\mu_{\rm t,b}$ of beam reselections as 
\begin{align}
\mu_{\rm t,b}(n) = \frac{2^n\sqrt{\lambda}}{\pi}v.
\end{align}

\end{document}